\documentclass[letterpaper, 10 pt, conference]{ieeeconf}
\IEEEoverridecommandlockouts
\usepackage{cite}
\usepackage{amsmath,amssymb,amsfonts}
\usepackage{algorithmic}
\usepackage{graphicx}
\usepackage{textcomp}
\usepackage{xcolor}
\usepackage{stfloats}
\usepackage{upgreek}
\usepackage{booktabs}
\usepackage{setspace}
\usepackage{multirow}
\usepackage{makecell}

\def\BibTeX{{\rm B\kern-.05em{\sc i\kern-.025em b}\kern-.08em
    T\kern-.1667em\lower.7ex\hbox{E}\kern-.125emX}}
    
\begin{document}

\title{MMA-Net: Multiple Morphology-Aware Network for Automated Cobb Angle Measurement\\
\thanks{This work is supported by National Natural Science Foundation of China grant \#62103181 and Shenzhen Outstanding Scientific and Technological Innovation Talents Training Project under Grant RCBS20221008093305007.}
\thanks{
Z. Qiu, J. Yang and J. Wang are with Shenzhen Key Laboratory of Robotics Perception and Intelligence and the Department of Electronic and Electrical Engineering at Southern University of Science and Technology in Shenzhen, China (e-mail: qiuzx2022@mail.sustech.edu.cn; yangj2021@mail.sustech.edu.cn; wangjk@sustech.edu.cn).}
\thanks{J. Wang is also with the Jiaxing Research Institute, Southern University of Science and Technology, Jiaxing, China.}
\thanks{* The authors contribute equally to this paper.}
\thanks{\# Corresponding author.}
}

\author{Zhengxuan Qiu*, Jie Yang*, $\text{Jiankun Wang}^{\#}$, \textit{Senior Member, IEEE}
}

\maketitle

\begin{abstract}
Scoliosis diagnosis and assessment depend largely on the measurement of the Cobb angle in spine X-ray images. With the emergence of deep learning techniques that employ landmark detection, tilt prediction, and spine segmentation, automated Cobb angle measurement has become increasingly popular. However, these methods encounter difficulties such as high noise sensitivity, intricate computational procedures, and exclusive reliance on a single type of morphological information. In this paper, we introduce the Multiple Morphology-Aware Network (MMA-Net), a novel framework that improves Cobb angle measurement accuracy by integrating multiple spine morphology as attention information. In the MMA-Net, we first feed spine X-ray images into the segmentation network to produce multiple morphological information (spine region, centerline, and boundary) and then concatenate the original X-ray image with the resulting segmentation maps as input for the regression module to perform precise Cobb angle measurement. Furthermore, we devise joint loss functions for our segmentation and regression network training, respectively. We evaluate our method on the AASCE challenge dataset and achieve superior performance with the SMAPE of 7.28\% and the MAE of 3.18°, indicating a strong competitiveness compared to other outstanding methods. Consequently, we can offer clinicians automated, efficient, and reliable Cobb angle measurement.
\end{abstract}

\section{Introduction}
Scoliosis is a prevalent spine condition characterized by abnormal lateral curvature and rotational deformities in the spine vertebra. In normal spine appearance from the front and back, the spine should be straight and centered over the pelvis while a spine with scoliosis exhibits a C- or S-shaped curve. Scoliosis affects 1\% to 4\% adolescents during pre-puberty growth\cite{cheng2015adolescent}. If left untreated, scoliosis can not only compromise patients’ physical appearance but also impair their psychological well-being and cardiopulmonary function \cite{barton2018adolescent}. Therefore, scoliosis assessment is crucial for adolescents.

\begin{figure}[t]
\centerline{\includegraphics[width=0.55\linewidth]{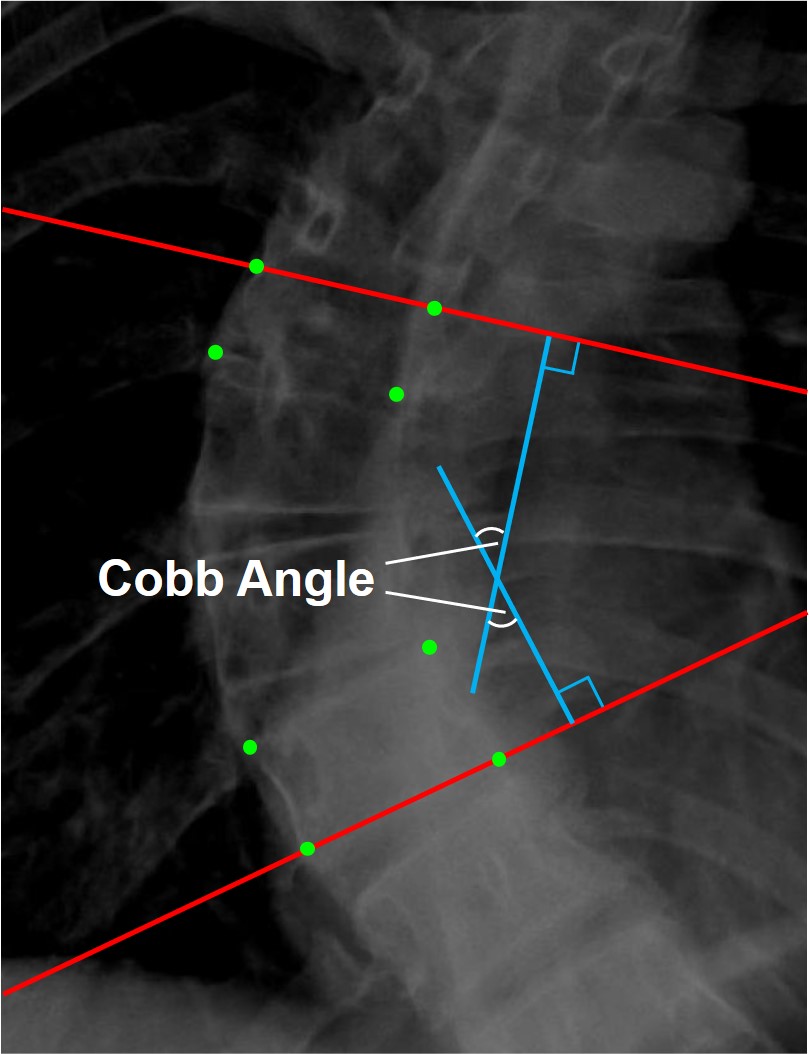}}
\caption{An example of Cobb angle measurement in an X-ray image.}
\label{fig1}
\vspace{-5mm}
\end{figure}

Clinicians confirm the diagnosis and assessment of scoliosis by X-ray images and a subsequent Cobb angle analysis of the images. The Cobb angle serves as a quantitative criterion for assessing the severity of scoliosis, which is defined as the angle between the line that borders the upper endplate (at the top of the superior end vertebra) and the line that borders the lower endplate (at the bottom of the inferior end vertebra), as shown in Fig.~\ref{fig1}. In current clinical practice, clinicians measure the Cobb angle manually. However, this manual measurement depends on experienced clinicians carefully identifying each vertebra and measuring the angles on X-ray images, which is time-consuming. Moreover, it is prone to subjective biases, resulting in potential variations of 5-10° in Cobb angle measurement among different clinicians \cite{wills2007comparison}. Hence, there is a need for an accurate and objective method for automated quantitative measurement of the Cobb angle.

With the advancement of computational technologies, deep learning methods are widely applied in medical image processing. In recent years, various studies based on deep learning have been conducted for automated measurement of the Cobb angle. According to the network prediction target type, these methods can be classified into three categories: landmark-based method, tilt-based method, and segmentation-based method.

For the landmark-based method, the network detects four landmarks of each vertebra to obtain its inclination and then calculates the Cobb angle by rules. Wu et al. \cite{wu2017automatic} proposed BoostNet, a CNN architecture that improves landmark detection accuracy by removing outlier features and minimizing intra-class variance. Chen et al. \cite{chen2020accurate} used RetinaNet to detect vertebrae and HR-Net to identify four landmarks within the vertebra area for more precise landmark detection. However, the quality of angle measurements relies heavily on the accuracy of landmark detection, and even small coordinates prediction errors can cause significant deviations in Cobb angle measurement. Due to the low contrast of X-ray images, it is common for the vertebral landmarks to be partially occluded, making errors in landmark detection inevitable. 

For the tilt-based method, the network predicts the tilt vector of each vertebra, which represents the inclination. Kim et al. \cite{kim2020automation} proposed a framework with two neural networks: Centroid-net for predicting vertebra centroids and M-net for predicting vertebral-tilt vectors. Similarly, Zou et al. \cite{zou2023vltenet} introduced VLTENet, a network for vertebra localization and tilt estimation, which improved Cobb angle measurement accuracy through fusion channel attention modules and a joint spine loss function. These methods essentially predict the tilt vector based on landmarks to estimate vertebra inclination, facing similar challenges as landmark-based methods. Moreover, the computational workflow of these methods is complex and intricate.

For the segmentation-based method, the network segments the vertebrae to improve Cobb angle measurement by extracting and utilizing spine morphological information more effectively. Wang et al. \cite{wang2020spinal} segmented two spine boundaries and combined them with the X-ray image to accurately predict Cobb angle using another network. Lin et al. \cite{lin2021seg4reg+} proposed Seg4Reg+, which combines spine region segmentation with Cobb angle regression using a triangle consistency learning scheme. These methods leverage the implicit relationship between the overall spine shape and the Cobb angle to enhance measurement accuracy. However, these methods only utilize a single morphological information of the spine, leading to insufficient learning of spine morphology.

In this paper, we present a novel deep-learning framework called MMA-Net, for automated Cobb angle measurement in X-ray images. The framework consists of a multiple morphology segmentation network and a multiple morphology-aware Cobb angle regression network. To leverage the three types of morphological information (spine region, centerline, and boundary), we first design a segmentation network that generates multiple morphological segmentation maps mentioned above. Then, we combine the obtained three segmentation maps with the raw X-ray image and feed them into a regression network to predict Cobb angles accurately. In contrast to solely using a single type of morphology as auxiliary information, we believe that providing multiple morphological information may better facilitate the network in learning the coarse-to-fine morphology of the spine. We summarize our main contributions as follows:

\begin{itemize}
    \item We propose the novel MMA-Net deep-learning framework, which exploits three types of spine morphological information to achieve Cobb angle measurement with high efficiency and precision. This auxiliary information can help the network focus on the proper spine area. 
    \item We develop a joint segmentation loss function that provides enhanced supervision for both region and edge segmentation. We also propose a joint regression loss function that is more task-specific and effectively optimizes the Cobb angle regression. 
    \item We evaluate the effectiveness of our proposed MMA-Net on a public dataset sourced from the AASCE Challenge \cite{wu2017automatic}. Extensive experiments demonstrate that our method outperforms the state-of-the-art (SOTA) methods for Cobb angle measurement. 
\end{itemize}

% The remainder of this paper is organized as follows. In Section II, we introduce the details of our proposed method, including the complete MMA-Net and loss functions. Section III reports the experimental results and ablation studies. Finally, we conclude our work in Section IV. 

\section{Methods}
\begin{figure*}[ht]
\centerline{\includegraphics[width=0.75\linewidth]{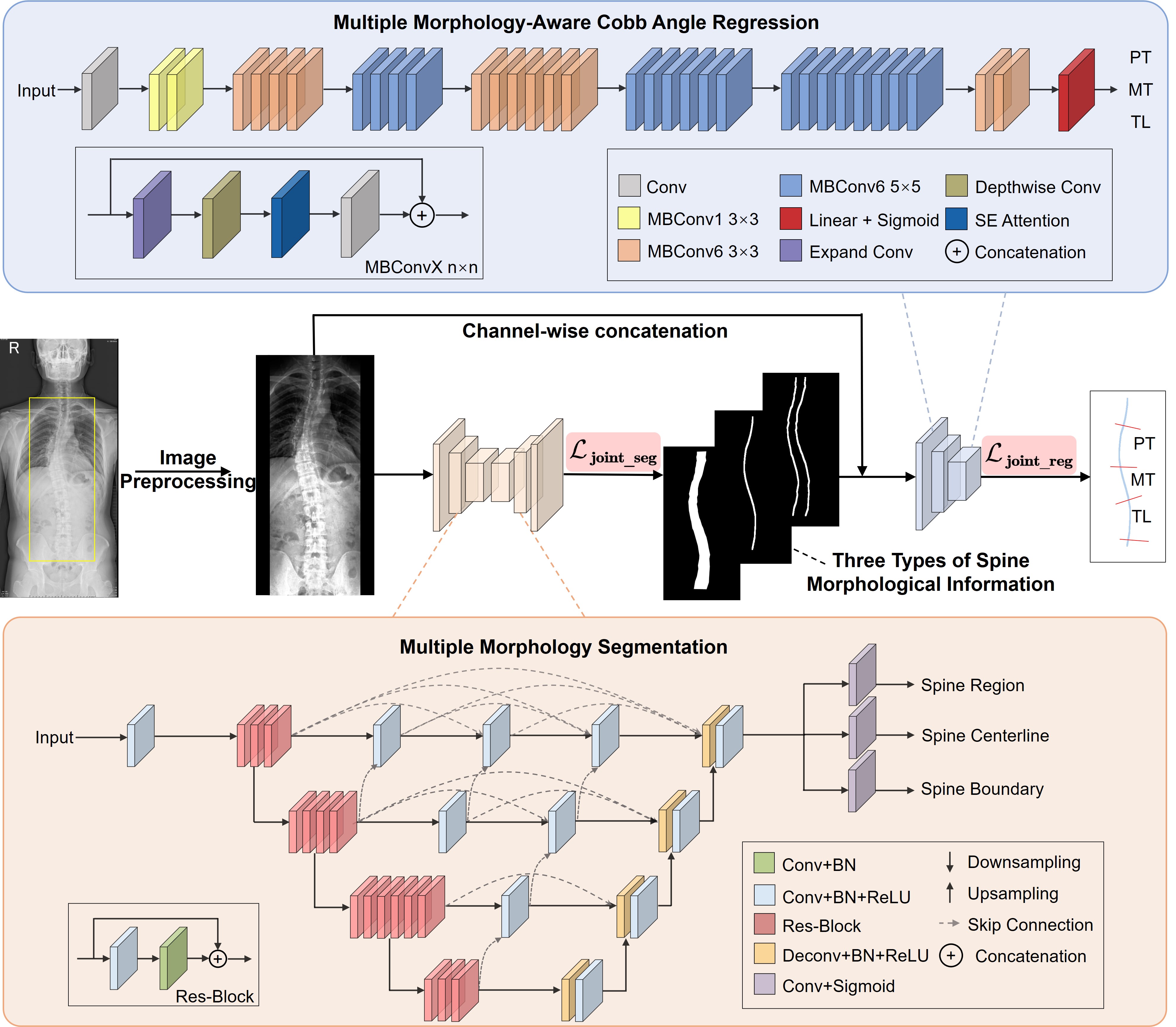}}
\caption{Overview of our proposed MMA-Net. First, the raw X-ray image undergoes image preprocessing (crop, resize, histogram equalization) before entering the next parts. The multiple morphology segmentation network, Res-UNet++, takes the preprocessed X-ray image as input and generates three types of morphological information (spine region, centerline, boundary). The multiple morphology-aware Cobb angle regression network, based on EfficientNet-b4, concatenates the X-ray image with the segmentation maps and predicts three Cobb angles. In the figure, MBConvX n$\times$n represents an expansion ratio of X and a filter size of n in the MBConv block.}
\label{fig2}
\vspace{-3mm}
\end{figure*}

The pipeline of MMA-Net is illustrated in Fig.~\ref{fig2}. The framework consists of three parts: multiple morphology segmentation, multiple morphology-aware Cobb angle regression, and joint loss functions. The segmentation network takes a preprocessed raw X-ray image as input and generates three types of morphological information (spine region, centerline, and boundary). After performing a channel-wise concatenation of the X-ray image and the obtained segmentation maps, we feed them into the regression network for accurately predicting the Cobb angle. 

\subsection{Multiple Morphology Segmentation}
The main idea of MMA-Net is to leverage the three types of spine multiple morphology to help the network achieve Cobb angle measurement with high efficiency and precision. Specifically, the spine multiple morphology is defined as follows: the spine region is the area enclosed by all vertebra landmarks, representing the overall morphological shape of the spine. The spine centerline is created by the center points based on the four corner landmark coordinates, indicating the vertebra center in the detailed spine curvature. The spine boundary is composed of two continuous curves formed by connecting landmarks on the left and right sides, representing the spinal and vertebral edge shape information. This multiple morphology can assist the Cobb regression network by providing auxiliary information from the overall spine area to the detailed vertebra center and edge information, from coarse to fine, thus increasing the accuracy of the Cobb angle measurement. From this point of view, we first design a multiple morphology segmentation network Res-UNet++ based on ResNet\cite{he2016deep} and UNet++\cite{zhou2018unet++}.  

The Res-UNet++ architecture consists of three main stages: encoder (feature extraction), decoder (feature reconstruction), and skip connections (feature fusion). To address network degradation caused by gradient disappearance and network depth, we utilize a reshaped ResNet34 in the encoder stage of Res-UNet++. The reshaped ResNet34 is primarily composed of res-blocks, each consisting of two 3$\times$3 convolutions, batch normalization, and ReLU activation. Depending on the kernel size and number of channels, four types of res-blocks are repeated 3, 4, 6, and 3 times, respectively, forming a 34-layer structure. The resulting feature vectors are then flattened and passed through the decoder section. The decoder stage involves upsampling modules. Each module comprises a 3$\times$3 deconvolution with a stride of 2, followed by a 3$\times$3 convolution, batch normalization, and ReLU activation. The output layer undergoes a 1$\times$1 convolution and uses a Sigmoid layer for normalization. This process helps determine whether each pixel is part of the segmentation target, producing the final result. In deep learning-based feature extraction, shallow features are simple and specific, capturing low-level details, while deep features tend to be complex and abstract, capturing high-level semantic information. Similar to UNet++, Res-UNet++ employs skip connections on dense convolutional blocks to fuse the shallow and deep features.

In general, the segmentation network Res-UNet++ takes the preprocessed raw X-ray image as input and outputs three types of morphological segmentation maps, laying the foundation for the latter Cobb angle measurement.

\subsection{Multiple Morphology-Aware Cobb Angle Regression}
The Cobb angle is typically calculated based on the morphological information of the spine. However, existing methods only use a single type of morphological information to assist Cobb angle measurement. By incorporating the spine multiple morphology as attention, the MMA-Net prioritizes the proper spine area and extracts valuable features, thereby boosting the accuracy of Cobb angle measurement. 

EfficientNet \cite{tan2019efficientnet} achieves superior results by simultaneously considering input resolution scaling, network depth, and width. For the Cobb angle regression network, we adopt EfficientNet-b4 based on its best performance among the EfficientNet family (EfficientNet b0-b7). EfficientNet-b4 comprises mobile inverted bottleneck convolution (MBConv) blocks, convolutional layers, a global average pooling layer, and a fully connected layer. Each MBConv block contains expansion, depthwise convolution, squeeze-and-excitation (SE) attention mechanism, and a convolutional layer. In the expansion stage, the number of channels in the layer is increased to make features wider. After expansion, depthwise convolution is conducted using a kernel size of 3$\times$3 or 5$\times$5. SE attention mechanism applies global average pooling to extract global features and squeeze along the channel dimension. Finally, we replace the last convolution layer with the output channel corresponding to three Cobb angles. 

In summary, we propose our multiple morphology-aware Cobb angle regression network based on EfficientNet-b4. The network takes the raw X-ray image concatenated with the spine region, centerline, and boundary as input and outputs three Cobb angles. 

\subsection{Joint Segmentation Loss and Joint Regression Loss}
Upon observing the X-ray images, it is evident that the regions of interest (especially the spine centerline and boundary) occupy a relatively small portion of the entire image. To enhance the region similarity between prediction maps and ground truth, we employ dice loss ($\mathcal{L}_{\text{DSC}}$) \cite{milletari2016v} based on dice similarity coefficient (DSC) to supervise the training of the segmentation network:
\begin{equation}
    \text{DSC} =  \frac{2 \sum_{{\it{i}}=1}^{\it{N}} \it{G_{i}}\cdot \it{P_{i}}}{\sum_{{\it i}=1}^{\it{N}} (\it{G_{i}} + \it{P_{i}})}, \enspace \mathcal{L}_{\text{DSC}} = 1 - \text{DSC} 
\end{equation}
where $N$ is the number of samples, $G_{i}$ and $P_{i}$ is the ground truth and the prediction of the $i$th sample. 

The edge of the segmentation maps can directly affect the quality of the entire area. However, commonly used loss functions like cross-entropy, IOU, and $\mathcal{L}_{\text{DSC}}$ tend to bias towards either the foreground or background, failing to minimize the edge information adequately. To alleviate this problem, we adopt boundary loss \cite{bokhovkin2019boundary} to enhance the edge details in the three spine segmentation maps. To construct boundary loss, we use the maximum pooling layer to extract the edges of the segmentation maps:
\begin{gather}
    G^{e}_{i} = \textbf{Maxpool}(1-G_i,\theta_0) - (1 - G_i)\\
    P^{e}_{i} = \textbf{Maxpool}(1-P_i,\theta_0) - (1 - P_i)\\
    G^{e,ext}_{i} = \textbf{Maxpool}(G^{e}_{i},\theta_1) \\
    P^{e,ext}_{i} = \textbf{Maxpool}(P^{e}_{i},\theta_1)  
\end{gather}
where $G^{e}_{i}$ and $P^{e}_{i}$ is the edge of ground truth and prediction of the $i$th sample. $G^{e,ext}_{i}$ and $P^{e,ext}_{i}$ denote the expanded edge. $\textbf{Maxpool}$ applies a pixel-wise maximum pooling operation to the input with a sliding window of size $\theta_0$ or $\theta_1$ (set to 3 and 5, respectively). After that precision and recall can be calculated. Then, the boundary metric ($\text{BF}_1$) and boundary loss ($\mathcal{L}_{\text{BF}_1}$) are defined as:
\begin{gather}
    \text{Pre} =  \frac{\sum_{i=1}^{N} P^{e}_{i} \cdot G^{e,ext}_{i}}{\sum_{i=1}^{N} P^{e}_{i}}, \enspace \text{Rec} =  \frac{\sum_{i=1}^{N} G^{e}_{i}\cdot P^{e,ext}_{i}}{\sum_{i=1}^{N} G^{e}_{i}}\\
    \text{BF}_1 = \frac{2 \text{Pre}\cdot \text{Rec}}{\text{Pre} + \text{Rec}}, \enspace \mathcal{L}_{\text{BF}_1} = 1 - \text{BF}_1
\end{gather}
Finally, our joint segmentation loss $\mathcal{L}_{\text{joint\_seg}}$ combines $\mathcal{L}_{\text{DSC}}$ and $\mathcal{L}_{\text{BF}_1}$:
\begin{equation}
    \mathcal{L}_{\text{joint\_seg}} = \mathcal{L}_{\text{DSC}} + \mathcal{L}_{\text{BF}_1}
\end{equation}

In order to minimize the error between the predicted angle and the ground truth in the Cobb angle measurement task, we design a joint regression loss function $\mathcal{L}_\text{joint\_reg}$. Serving as the final evaluation metric in the AASCE Challenge, symmetric mean absolute percentage error (SMAPE) is an accuracy measure based on relative errors. In addition, considering that mean absolute error (MAE) is a metric describing the absolute error, we add it to our joint regression loss. To further minimize the angle error, we incorporate the circular mean absolute error (CMAE), which is specifically designed for angle-like quantities.  
\begin{equation}
\label{eq9}
    \text{SMAPE} = \mathcal{L}_{\text{SMAPE}} = \frac{1}{N} \sum_{i=1}^N \frac{\sum_{j=1}^3 |G_{ij} - P_{ij}|}{\sum_{j=1}^3 |G_{ij} + P_{ij} + \varepsilon|} \times 100\%
\end{equation}

\vspace{-2mm}

\begin{equation}
\label{eq10}
    \text{MAE} = \mathcal{L}_{\text{MAE}} = \frac{1}{N\cdot 3} \sum_{i=1}^N \sum_{j=1}^3 |G_{ij} - P_{ij}|
\end{equation}

\begin{equation}
\label{eq11}
    \text{CMAE} = \mathcal{L}_{\text{CMAE}} = \frac{1}{N} \sum_{{i}=1}^{N} \textbf{arctan} \frac{\sum_{{j}=1}^{3} \textbf{sin}(G_{{ij}} - P_{{ij}})}{\sum_{{j}=1}^{3} \textbf{cos}(G_{{ij}} - P_{{ij}})}
\end{equation}

\vspace{-4mm}

\begin{equation}
\label{eq12}
    \mathcal{L}_\text{joint\_reg} = \mathcal{L}_{\text{SMAPE}} + \mathcal{L}_{\text{MAE}} + \mathcal{L}_{\text{CMAE}}
\end{equation}
where $G_{{ij}}$ and $P_{{ij}}$ denote the ground truth and the prediction of the $j$th angle of the $i$th sample. $\varepsilon$ is the smooth factor (set to $\text{1}\times\text{10}^{\text{-30}}$).

\section{Experiments}
\subsection{Datasets}
The dataset used for evaluation is the AASCE Challenge dataset, which comprises 609 spine (AP) X-ray images. The dataset is divided by the provider into 481 images for training and 128 images for testing. Each image in the dataset contains 17 vertebrae, and each vertebra has been manually annotated by experienced clinicians using four landmarks located in the corners. The three Cobb angles, proximal thoracic (PT), main thoracic (MT), and thoracolumbar/lumbar (TL) are derived from these landmarks.

\subsection{Implementation Details}
Based on the definition of spine multiple morphology mentioned above in Section 2. A, we generate the ground truth for multiple morphology segmentation. Besides, we enhance the clarity of the spine centerline and boundary by applying a dilation operation with a kernel size of 5.  To standardize the input size for the network, the images are cropped and resized to a fixed size [512, 256]. To address the domain gap between training and testing sets, histogram equalization is applied to both sets to enhance their similarity. Then we augment the dataset by flipping, rotating [-25°-25°], and scaling with the random factor between [0.85, 1.25] to alleviate overfitting during training. 

\begin{table*}[ht]
\setlength{\abovecaptionskip}{0.05cm} %设置三线表标题与第一条线间距
\centering
\caption{Comparison of spine region, centerline, and boundary segmentation performance with different methods}
\setlength{\tabcolsep}{4mm}
\begin{tabular}{ccccccc}
\toprule[1.0pt]
\multirow{2}{*}{Method} & \multicolumn{2}{c}{Region}    & \multicolumn{2}{c}{Centerline} & \multicolumn{2}{c}{Boundary}  \\  \cmidrule(r){2-3} \cmidrule(r){4-5} \cmidrule(r){6-7}
           & DSC [\%]  & $\text{BF}_\text{1}$ [\%]  & DSC [\%] & $\text{BF}_\text{1}$ [\%]  & DSC [\%]  & $\text{BF}_\text{1}$ [\%]  \\ \hline
UNet \cite{ronneberger2015u}       & 90.2 & 66.0 & 65.8 & 76.4 & 80.4 & 88.0 \\
PSPNet\cite{zhao2017pyramid}     & 89.7 & 69.6 & 67.3 & 77.4 & 78.5 & 85.9 \\
Res-UNet\cite{zhang2018road}   & 91.6 & 72.3 & 69.8 & 78.9 & 80.3 & 87.7 \\
UCTransNet\cite{wang2022uctransnet} & 91.1 & 72.0 & 70.8 & 80.3 & 81.3 & 88.7 \\
Res-UNet++ ($\mathcal{L}_{\text{DSC}}$)           & 91.6 & 70.6 & 71.1 & 78.0 & 81.7 & 88.3 \\
Res-UNet++ ($\mathcal{L}_{\text{joint\_seg}}$)    & \textbf{91.8} & \textbf{74.3} & \textbf{71.6}  & \textbf{81.4} & \textbf{82.0} & \textbf{89.7} \\
\bottomrule[1.0pt]
\label{tabel1}
\end{tabular}
\vspace{-2mm}
\end{table*}

\begin{table*}[ht]
\vspace{-2mm}
\setlength{\abovecaptionskip}{0.05cm} %设置三线表标题与第一条线间距
\centering
\caption{Comparison of Cobb angle measurement performance with SOTAs}
\setlength{\tabcolsep}{4mm}
\begin{tabular}{@{}ccccccc@{}}
\toprule[1.0pt]
Method &
  \begin{tabular}[c]{@{}c@{}}MAE {[}°{]}\end{tabular} &
  \begin{tabular}[c]{@{}c@{}}SMAPE {[}\%{]}\end{tabular} &
  \begin{tabular}[c]{@{}c@{}}CMAE {[}°{]}\end{tabular} &
  \begin{tabular}[c]{@{}c@{}}ED {[}°{]}\end{tabular} &
  \begin{tabular}[c]{@{}c@{}}MD {[}°{]}\end{tabular} &
  \begin{tabular}[c]{@{}c@{}}CD {[}°{]}\end{tabular} \\ \midrule
 SCG-Net\cite{wang2020spinal}   & -             & 22.18         & 4.91          & 11.23         & 14.74          & 10.17         \\
 Seg4Reg\cite{lin2020seg4reg}   & -             & 21.71         & 4.85          & 11.17         & 14.55          & 10.16         \\
 KEF\cite{guo2021heterogeneous} & -             & 8.62          & -             & -             & -              & -             \\
 VF\cite{kim2020automation}     & 3.51          & 7.84          & -             & -             & -              & -             \\
 Seg4Reg+\cite{lin2021seg4reg+} & 3.73          & 7.32          & -             & -             & -              & -             \\
 MMA-Net (Ours)                 & \textbf{3.18} & \textbf{7.28} & \textbf{2.26} & \textbf{6.59} & \textbf{9.56} & \textbf{5.68} \\ \bottomrule[1.0pt]
\end{tabular}
\vspace{-5mm}
\end{table*}

The MMA-Net is implemented using Pytorch on a single NVIDIA RTX 4090. We train the segmentation network for 400 epochs using Adam optimization with learning rate $\text{1}\times\text{10}^{\text{-4}}$ and cosine decay schedule. The batch size is set to 16. For the Cobb angle regression network, we train the network for 1200 epochs with a learning rate $\text{1}\times\text{10}^{\text{-5}}$ and cosine decay schedule, and the batch size remains at 16.

\subsection{Evaluation Metrics}
To assess the performance of the segmentation network, we employ two evaluation metrics DSC and $\text{BF}_1$. The definitions and descriptions of these metrics can be found in Equation (1) and (7).

To evaluate the performance of the Cobb angle regression network, we use five evaluation metrics including MAE, SMAPE, CMAE [their definition can be found in Equation (9-11)], Euclidean distance (ED), Manhattan distance (MD), and Chebyshev distance (CD).

\begin{equation}
\label{eq13}
    \text{ED} = \frac{1}{N} \sum_{{i}=1}^{N} \sqrt{\sum_{{j}=1}^{3} (G_{{ij}} - P_{{ij}})^2}
\end{equation}

\begin{equation}
\label{eq14}
    \text{MD} = \frac{1}{N} \sum_{{i}=1}^{N} \sum_{{j}=1}^{3} |G_{{ij}} - P_{{ij}}|
\end{equation}

\vspace{-2mm}

\begin{equation}
\label{eq15}
    \text{CD} = \frac{1}{N} \sum_{{i}=1}^{N} \textbf{max}(|G_{{i} 1} - P_{{i} 1}|, |G_{{i} 2} - P_{{i} 2}|, |G_{{i} 3} - P_{{i} 3}|)
\end{equation}

\subsection{Comparison with SOTA Methods}

We compare our segmentation network Res-UNet++ with other SOTA methods including popular medical segmentation baselines (i.e., UNet \cite{ronneberger2015u}, ResUNet \cite{zhang2018road} and UCTransNet \cite{wang2022uctransnet}) as well as the semantic segmentation baseline (i.e., PSPNet \cite{zhao2017pyramid}, which performs superior performance in the spine segmentation task \cite{lin2021seg4reg+}\cite{lin2020seg4reg}). All models are trained with the same experiment settings for a fair comparison. Table I lists the segmentation performance of the above-mentioned methods, evaluated by DSC and $\text{BF}_\text{1}$. We can observe that Res-UNet++ achieves the best performance across all three segmentation tasks (spine region, centerline, and boundary). For instance, Res-UNet++ gets the highest DSC of 91.8\% and the highest $\text{BF}_1$ of 74.3\% in the spine region segmentation task. This can potentially be attributed to Res-UNet++’s strong ability for multi-scale feature integration. 

\begin{figure}[ht]
\vspace{-2mm}
\centerline{\includegraphics[width=1.0\linewidth]{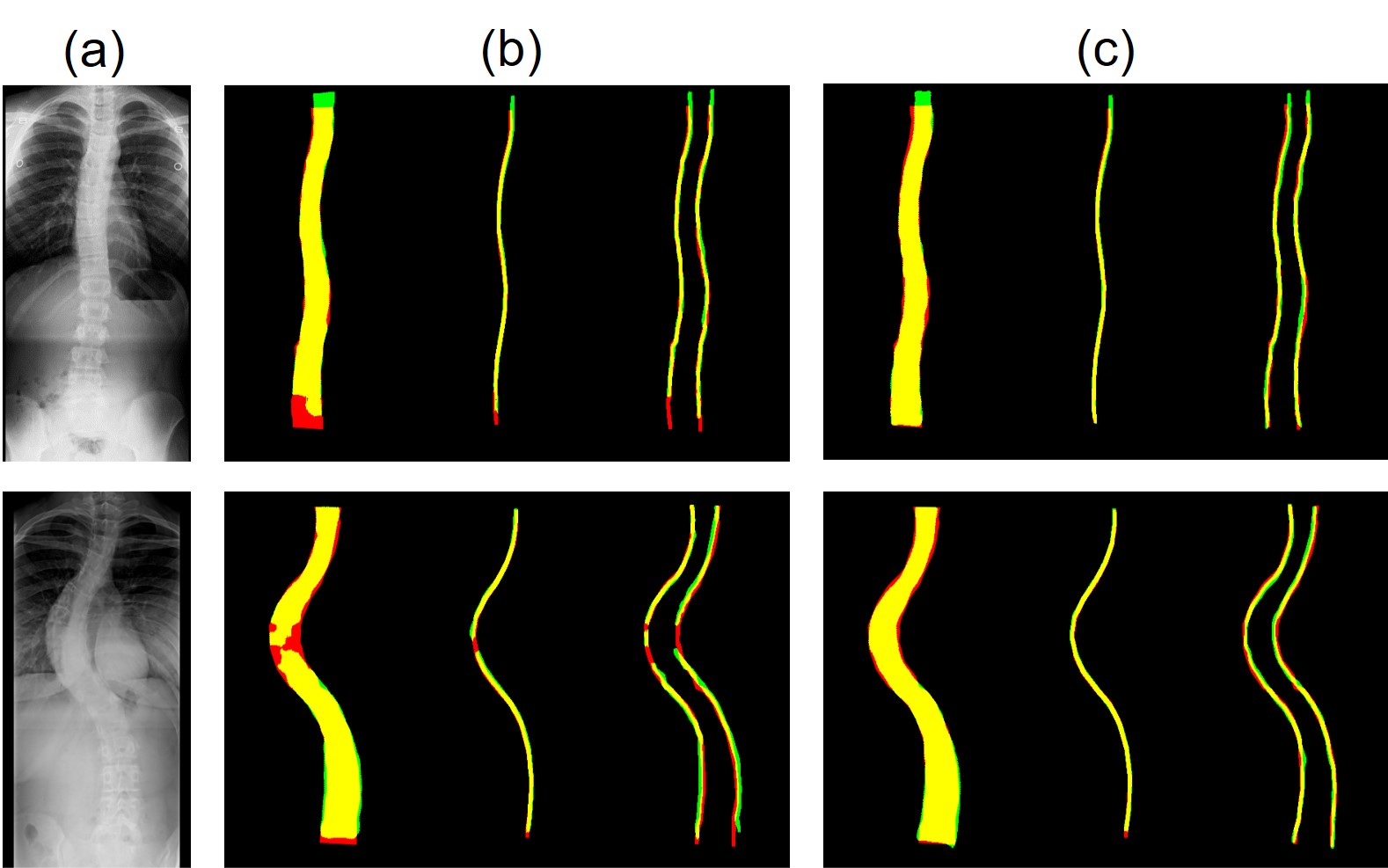}}
\caption{Examples of multiple morphological information segmentation (spine region, centerline, and boundary). Each group shows (a) the raw image, (b) the segmentation results with $\mathcal{L}_{\text{DSC}}$, and (c) the segmentation results with $\mathcal{L}_{\text{joint\_seg}}$, respectively. The yellow mask is true positive, the red mask is false negative, and the green mask is false positive.}
\label{fig3}
\vspace{-6mm}
\end{figure}

A series of experiments are conducted to compare our proposed MMA-Net framework with other SOTA methods on the AASCE challenge dataset. To better validate the effectiveness of MMA-Net, we choose three different categories of deep learning-based Cobb angle measurement methods, including landmark-based method (KEF\cite{guo2021heterogeneous}), tilt-based method (VF\cite{kim2020automation}), and segmentation-based method (SCG-Net\cite{wang2020spinal}, Seg4Reg\cite{lin2020seg4reg}, and Seg4Reg+\cite{lin2021seg4reg+}). TABLE II demonstrates that our proposed framework exhibits superior performance advantages over the other five competing methods. Specifically, our proposed framework MMA-Net not only achieves the lowest MAE, SMAPE, and CMAE metric values of 3.18°, 7.28\%, and 2.26°, respectively, but also attains the lowest values in other metrics including ED, MD, and CD. Notably, the error range for manual Cobb angle measurement by clinicians is 5-10° \cite{wills2007comparison}. Surprisingly, MMA-Net exhibits an average error (MAE and CMAE) of less than 3° in predicting the Cobb angle. Therefore, in a clinical diagnostic setting, our proposed framework has the potential to effectively assist clinicians in assessing scoliosis conditions.

\subsection{Ablation Analysis}

To verify the effectiveness of our joint segmentation loss function in Res-UNet++, we conduct an ablation experiment. We compare ResUNet++ trained with dice loss [Res-UNet++ ($\mathcal{L}_{\text{DSC}}$)] and our joint segmentation loss [Res-UNet++($\mathcal{L}_{\text{joint\_seg}}$)]. TABLE I demonstrates that Res-UNet++ ($\mathcal{L}_{\text{joint\_seg}}$) shows a distinct advantage as it achieves a higher DSC and $\rm BF1$ value in all three segmentation tasks compared to Res-UNet++ ($\mathcal{L}_{\text{DSC}}$). Furthermore, Fig.~\ref{fig3} visualizes and compares the segmentation results of Res-UNet++ using different loss functions. By incorporating boundary loss, the network can prioritize the edge information, facilitating the segmentation of the overall area in the spine region task. The spine centerline and boundary segmentation results demonstrate that Res-UNet++ ($\mathcal{L}_{\text{joint\_seg}}$) preserves connectivity with less deviation from the ground truth. Overall, our joint segmentation loss enhances the network's ability to accurately segment the spine region, centerline, and boundary.

\begin{table*}[ht]
\setlength{\abovecaptionskip}{0.05cm} %设置三线表标题与第一条线间距
\centering
\caption{Ablation study results of input and loss function for Cobb angle measurement}
\begin{tabular}{@{}ccccccccccccc@{}}
\toprule
\multicolumn{4}{c}{Input components} &
  \multicolumn{3}{c}{Loss} &
  \multicolumn{1}{c}{\multirow{2}{*}{\makecell{MAE\\{[}°{]}}}} &
  \multicolumn{1}{c}{\multirow{2}{*}{\makecell{SMAPE\\{[}\%{]}}}} &
  \multicolumn{1}{c}{\multirow{2}{*}{\makecell{CMAE\\{[}°{]}}}} &
  \multicolumn{1}{c}{\multirow{2}{*}{\makecell{ED\\{[}°{]}}}} &
  \multicolumn{1}{c}{\multirow{2}{*}{\makecell{MD\\{[}°{]}}}} &
  \multicolumn{1}{c}{\multirow{2}{*}{\makecell{CD\\{[}°{]}}}} \\ \cmidrule(r){1-4} \cmidrule(r){5-7}
  Image &
  Region &
  Centerline &
  Boundary &
  $\mathcal{L}_{\text{SMAPE}}$ &
  $\mathcal{L}_{\text{MAE}}$ &
  $\mathcal{L}_{\text{CMAE}}$ &
  \multicolumn{1}{c}{} &
  \multicolumn{1}{c}{} &
  \multicolumn{1}{c}{} &
  \multicolumn{1}{c}{} &
  \multicolumn{1}{c}{} &
  \multicolumn{1}{c}{} \\ \midrule
 \checkmark &  &  &  &\checkmark  &\checkmark  &\checkmark  &4.64  &10.34  &3.25  &9.24  &13.93  &7.63  \\
 \checkmark & \checkmark  &  &  &\checkmark  &\checkmark  &\checkmark  &3.64  &8.16  &2.66  &7.34  &10.92  &6.05  \\
 \checkmark &\checkmark  &\checkmark  &  &\checkmark  &\checkmark  &\checkmark  &3.43  &7.80  &2.51  &7.09  &10.30  &6.04  \\
 \checkmark &\checkmark  &\checkmark  &\checkmark  &\checkmark  &\checkmark  &\checkmark  &\textbf{3.18}  &\textbf{7.28}  &\textbf{2.26}  &\textbf{6.59}  &\textbf{9.56}  &\textbf{5.68}  \\
 \checkmark &\checkmark  &\checkmark  &\checkmark  &\checkmark  &  &  &3.70  &8.49  &2.56  &7.54  &11.10  &6.36  \\
 \checkmark &\checkmark  &\checkmark  &\checkmark  &\checkmark  &\checkmark  &  &3.49  &7.83  &2.45  &7.13  &10.47  &6.01  \\ \bottomrule
 \label{table3}
\end{tabular}
\vspace{-5mm}
\end{table*}

\begin{figure*}[ht]
\vspace{-3mm}
\centerline{\includegraphics[width=0.8\linewidth]{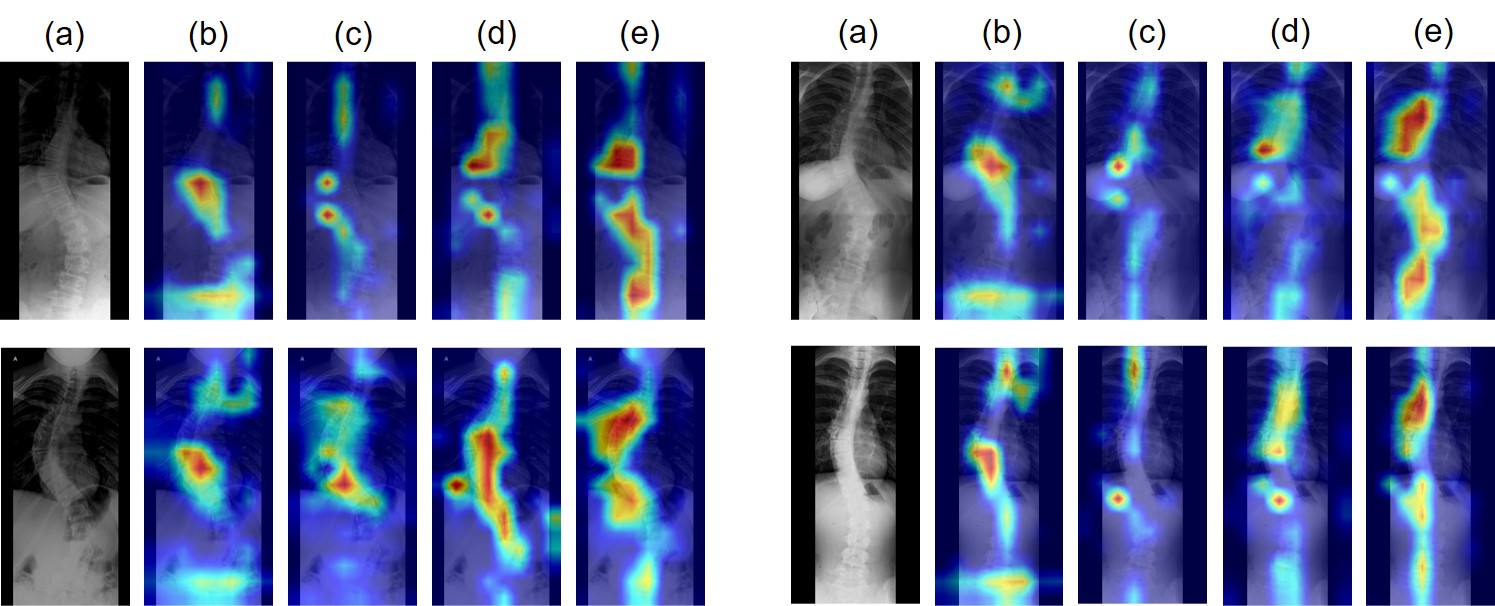}}
\caption{Representative examples of Grad-CAM heatmaps computed by MMA-Net with different input conditions. Each group consists of five images, from left to right, labeled as follows: (a) raw image; (b) input (image); (c) input (image + region); (d) input (image + region + centerline); (e) input (image + region + centerline + boundary).}
\label{fig4}
\vspace{-5mm}
\end{figure*}

We analyze the impact of input components and loss functions in our MMA-Net framework. In Table~\ref{table3}, we conduct ablation experiments by combining different input components and loss functions. As the spine morphological information is gradually incorporated, the regression network exhibits improved performance and generalizability in Cobb angle measurement. This improvement is attributed to the auxiliary morphological information provided by the spine region, centerline, and boundary, which allows the network to focus on learning from the proper spine area. Additionally, compared to using only $\mathcal{L}_\text{SMAPE}$, employing our proposed $\mathcal{L}_\text{joint\_reg}$ brings 1.21\% improvement in the SMAPE metric. The improvement can be explained as the unique design of combining the three different loss functions is more suitable for minimizing the error between predicted angles and ground truth. Fig.~\ref{fig4} illustrates the influence of different input conditions on MMA-Net using Grad-CAM heatmaps \cite{selvaraju2017grad}. The representative examples demonstrate that providing only raw images causes the network to not accurately focus on the most curved part of the spine, leading to greater measurement errors. By incorporating the spine region into the input, the network extends its attention to a wider scope within the spine area. Furthermore, introducing the spine centerline and boundary effectively constrains the network's attention to the entire spine area, thereby boosting the performance of Cobb angle measurement. In our framework, the combination of all input components and loss functions achieves the best performance, proving the necessity and effectiveness of 
each component and loss function. 

\vspace{0.01\textwidth}

\section{Conclusions and Future Work}
In this paper, we propose a novel deep-learning framework MMA-Net for accurate and automated Cobb angle measurement. The main idea is to incorporate multiple morphological information including spine region, centerline, and boundary to help the network focus on proper spine area. To better supervise the network training, we design joint loss functions for segmentation and regression networks, respectively. Experimental results on the AASCE challenge dataset demonstrate the superiority of MMA-Net. In the future, we will explore flexible network architectures that consider the complexity of clinical diagnosis and ensure accurate Cobb angle measurement for various clinical applications.

\clearpage
\bibliographystyle{IEEEtran}
\bibliography{ref}

\end{document}